\documentclass[useAMS,usenatbib,a4paper]{mn2e}

\usepackage{txfonts}
\usepackage{amssymb}
\usepackage{graphicx}
\usepackage{subfigure}

\usepackage{ifpdf}
\ifpdf
  \usepackage[pdftex]{hyperref}
\else
  \usepackage[ps2pdf,colorlinks=true]{hyperref}
\fi

\newcommand{\fig}[1]
	{Figure~#1}
\newcommand{\Fig}[1]
	{Figure~#1}
\newcommand{\sectn}[1]
	{Section~#1}

\newcommand{\appn}[1]
	{Appendix~#1}

\newcommand{\astroph}{\mbox{astro-ph}}

\newcommand{\mnras}{\mbox{MNRAS}}
\newcommand{\physd}{Phys. Rev. D.}
\newcommand{\physlett}{Phys. Rev. Lett.}
\newcommand{\apj}{ApJ}
\newcommand{\apjl}{ApJL}
\newcommand{\apjs}{ApJS}
\newcommand{\acha}{Applied and Computational Harmonic Analysis}
\newcommand{\jmp}{J.\ of Math.\ Phys.}

\newcommand{\cmb}
	{{CMB}}
\newcommand{\cmbtext}
	{cosmic microwave background}

\newcommand{\cswt}
	{{CSWT}}
\newcommand{\cswttext}
	{continuous spherical wavelet transform}

\newcommand{\morlet}
	{real Morlet}
\newcommand{\Morlet}
	{Real Morlet}

\newcommand{\wmap}
        {{WMAP}}
\newcommand{\wmaptext}
        {Wilkinson Microwave Anisotropy Probe}

\newcommand{\lambdaarch}
	{{LAMBDA}}

\newcommand{\kpzero}
	{{Kp0}}

\newcommand{\healpix}
	{{HEALPix}}

\newcommand{\etal}
	{\mbox{et al.}}

\newcommand{\ie}
	{\mbox{i.e.}}

\newcommand{\scale}
	{\ensuremath{a}}

\newcommand{\eulerc}
	{\ensuremath{\gamma}}

\newcommand{\spcend}
	{\ensuremath{\:}}

\newcommand{\ngsim}
	{1000}

\newcommand{\nstdmorskewteg}  
	{\mbox{$6.42$}}

\newcommand{\nstdmorskewtegsgn}  
	{\mbox{$-6.42$}}

\title[Non-Gaussianity in the \wmap\ 3-year data]
  {A high-significance detection of non-Gaussianity in the \wmap\ 3-year %
   data using directional spherical wavelets}

\author[J.~D.~McEwen \etal]
  {J.~D.~McEwen,$^1$\thanks{E-mail: mcewen@mrao.cam.ac.uk} 
   M.~P.~Hobson,$^1$ A.~N.~Lasenby$^1$ and D.~J.~Mortlock$^2$\\
  $^1$Astrophysics Group, 
      Cavendish Laboratory, J.~J.~Thomson Avenue,
      Cambridge CB3 0HE, UK\\
  $^2$Blackett Laboratory, Imperial College of Science, Technology and Medicine,
    Prince Consort Road, London SW7 2BW, UK}
\date{\today}
\pagerange{\pageref{firstpage}--\pageref{lastpage}} 
\pubyear{2006}

\def\LaTeX{L\kern-.36em\raise.3ex\hbox{a}\kern-.15em
    T\kern-.1667em\lower.7ex\hbox{E}\kern-.125emX}

\begin{document}
\label{firstpage}
\maketitle


\begin{abstract}
We repeat the directional spherical \morlet\ wavelet analysis, used to detect non-Gaussianity in the \wmaptext\ (\wmap) 1-year data \citep{mcewen:2005a}, on the \wmap\ 3-year data.  The non-Gaussian signal previously {detect\-ed} is indeed present in the 3-year data, although the significance of the detection is reduced.  Using our most conservative method for constructing significance measures, we find the {signi\-fi\-cance} of the detection of non-Gaussianity drops from $98.3\pm0.4$\% to $94.9\pm0.7$\%; the {signi\-fi\-cance} drops from $99.3\pm0.3$\% to $97.2\pm0.5$\% using a method based on the $\chi^2$ statistic.  The wavelet analysis allows us to localise most likely sources of non-Gaussianity on the sky.  We detect very similar localised regions in the \wmap\ 1-year and 3-year data, although the regions extracted appear more pronounced in the 3-year data.  When all localised regions are excluded from the analysis the 3-year data is consistent with \mbox{Gaussianity}. 
\end{abstract}


\begin{keywords}
 cosmic microwave background -- methods: data analysis -- methods: numerical
\end{keywords}


\section{Introduction}

Recent measurements of the \cmbtext\ (\cmb) anisotropies, in particular those made by the \wmaptext\ (\wmap), provide data of unprecedented precision with which to study the origin of the universe.  Such observations have lent strong support to the standard cosmological concordance model.  Nevertheless, many details and assumptions of the concordance model are still under close scrutiny.
One of the most important and topical assumptions of the standard model is that of the statistics of the primordial fluctuations that give rise to the anisotropies of the \cmb.  In the simplest inflationary models, primordial perturbations seed Gaussian temperature fluctuations in the \cmb\ that are statistically isotropic over the sky.  However, this is not necessarily the case for non-standard inflationary models or various cosmic defect scenarios.

The assumptions of Gaussianity and isotropy have been questioned recently with many works highlighting deviations from Gaussianity in the \wmap\ 1-year data (\wmap 1; \citealt{bennett:2003}), calculating measures such as
the bispectrum and Minkowski functionals
  \citep{komatsu:2003,mm:2004,lm:2004,medeiros:2005},
the genus
  \citep{cg:2003,eriksen:2004}, 
correlation functions
  \citep{gw:2003,eriksen:2005,tojeiro:2005},
low-multipole alignment statistics 
  \citep{oliveira:2004,copi:2004,copi:2005,schwarz:2004,slosar:2004,weeks:2004,lm:2005a,lm:2005b,lm:2005c,lm:2005d,bielewicz:2005,oliveira:2006}, 
structure alignment statistics 
  \citep{wiaux:2006}
phase associations
  \citep{chiang:2003,chiang:2004,coles:2004,dineen:2005},
local curvature 
  \citep{hansen:2004,cabella:2005},
the higher criticism statistic
  \citep{cayon:2005},
hot and cold spot statistics
  \citep{larson:2004,larson:2005},
fractal statistics
  \citep{sadegh:2006}
and wavelet coefficient statistics 
  \citep{vielva:2003,mw:2004,mcewen:2005a,mcewen:2005c,cruz:2005,cruz:2006a}.
Some statistics show consistency with Gaussianity, whereas others provide some evidence for a non-Gaussian signal and/or an asymmetry between the northern and southern Galactic hemispheres.  
Although the recently released \wmap\ 3-year data (\wmap3; \citealt{hinshaw:2006}) is consistent with the \wmap 1 data, a more thorough treatment of beams, foregrounds and systematics in the 3-year data, in addition to a further two years of observing time, mean that \wmap3 provides a more reliable data-set on which to confirm or \mbox{refute} previous results.
A Gaussianity analysis is performed on the \wmap3 data by \citet{spergel:2006}, using the one point distribution function, Minkowski functionals, the bispectrum and the trispectrum.  No evidence is found for non-Gaussianity; however, the authors do not re-evaluate the large number of statistical tests that have been used to detect non-Gaussianity in the \wmap1 data.
Indeed, deviations from Gaussianity and isotropy have recently been detected in the \wmap3 data, using measures such as
anisotropy statistics 
  \citep{helling:2006,bernui:2006}
phase associations
  \citep{chiang:2006}
and wavelet coefficient statistics 
  \citep{cruz:2006b}, 
with little change in the significance levels obtained by each technique for the first and third year data.
Although the departures from Gaussianity and isotropy detected in the \wmap1 and \wmap3 data may simply highlight unremoved foreground contamination or other systematics, which itself is of importance for cosmological inferences drawn from the data, if the source of these detections is of cosmological origin then this would have important implications for the standard cosmological model.

In this letter we focus on the significant detection of non-Gaussianity that we made previously in the \wmap1 data using directional spherical wavelets \citep{mcewen:2005a}, to see if the detection is still present in the \wmap3 data.
The remainder of this letter is organised as follows.  In \sectn{\ref{sec:analysis}} we briefly review the analysis procedure and discuss the data maps considered.  Results are presented and discussed in \sectn{\ref{sec:results}}, before concluding remarks are made in \sectn{\ref{sec:conclusions}}.


\section{Non-Gaussianity analysis}
\label{sec:analysis}

We repeat on the \wmap 3 data our non-Gaussianity analysis performed previously on the \wmap 1 data \citep{mcewen:2005a}, focusing only on the most significant detection of non-Gaussianity made previously.  We refer the reader to our previous work \citep{mcewen:2005a} for a detailed description of the analysis procedure and present here only a very brief overview.

We apply a spherical wavelet analysis to probe the \wmap\ data for non-Gaussianity.  Wavelets are an ideal tool to search for deviations from Gaussianity due to the scale and spatial localisation inherent in a wavelet analysis.  
To perform a wavelet analysis of full-sky \cmb\ maps we apply our fast \cswttext\ (\cswt; \citealt{mcewen:2005b}), which is based on the spherical wavelet transform developed by Antoine, Vandergheynst and colleagues \citep{antoine:1998,antoine:1999,antoine:2002,antoine:2004,wiaux:2005} and the fast spherical convolution developed by \citet{wandelt:2001}.  We use only the \morlet\ wavelet in this analysis since it gave the most significant detection of non-Gaussianity in the \wmap 1 data \citep{mcewen:2005a}.

To minimise the contribution of foregrounds and systematics to \cmb\ anisotropy measurements, the \wmap\ assembly contains a number of receivers that observe at a range of frequencies \citep{bennett:2003}.  In this analysis we consider the signal-to-noise ratio enhanced co-added map constructed from the \wmap 3 data.  This map is constructed by the same procedure described generally by \citet{komatsu:2003} and described in the context of our non-Gaussian analysis by \citet{mcewen:2005a}.
We use the foreground reduced sky maps and apply the \kpzero\ mask to remove residual Galactic emission and known point sources.  The foreground maps and mask are available from the Legacy Archive for Microwave Background Data Analysis (\lambdaarch) website\footnote{http://cmbdata.gsfc.nasa.gov/}.

To quantify the significance of any deviations from Gaussianity we perform \ngsim\ Monte Carlo simulations.  This involves simulating \ngsim\ Gaussian co-added maps.  Each simulated map is constructed in an analogous manner to the co-added map constructed from the data.  A Gaussian \cmb\ realisation is simulated from the theoretical power spectrum fitted by the \wmap\ team (the power spectrum we use is also available from \lambdaarch).
Measurements made by the various receivers are then simulated by convolving with realistic beams and adding anisotropic \wmap 3 noise for each receiver.  The simulated observations for each receiver are then combined to give a co-added map.

To probe the \wmap 3 data for deviations from Gaussianity the skewness of the \morlet\ wavelet coefficients is examined over a range of scales and orientations (the scales and orientations considered are defined in \citealt{mcewen:2005a}).  Any deviation from zero is an indication of non-Gaussianity in the data.  An identical analysis is performed on the \ngsim\ Gaussian simulations to quantify the significance of any deviations.


\section{Results}
\label{sec:results}

\begin{figure}
\centering
  \includegraphics[angle=-90,width=75mm]{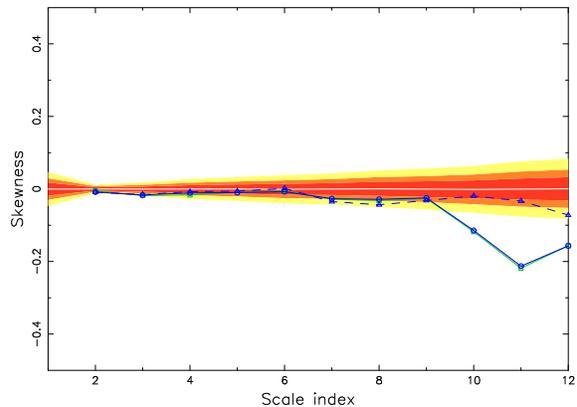}
\caption{\Morlet\ wavelet coefficient skewness statistics ($\eulerc=72^\circ$).  Points are plotted for the \wmap 1 data (solid, green, squares), \wmap 3 data (solid, blue, circles) and the \wmap 3 data with localised regions removed (dashed, blue, triangles).
Confidence regions obtained from \ngsim\ \wmap 3 Monte Carlo simulations are shown for 68\% (red), 95\%
  (orange) and 99\% (yellow) levels, as is the mean (solid white
  line).
}
\label{fig:stats}
\end{figure}

The skewness of the \morlet\ wavelet coefficients of the co-added \wmap 3 map are displayed in \fig{\ref{fig:stats}}, with confidence intervals constructed from the \ngsim\ \wmap 3 Monte Carlo simulations also shown.  Only the plot corresponding to the orientation of the maximum deviation from Gaussianity is shown.  The non-Gaussian signal present in the \wmap 1 data is clearly present in the \wmap 3 data.  In particular, the large deviation on scale $\scale_{11}=550\arcmin$ and orientation $\eulerc=72^\circ$ is almost identical (although it is in fact very marginally lower in the \wmap 3 data).

\begin{figure}
\centering
  \includegraphics[angle=-90,width=75mm]{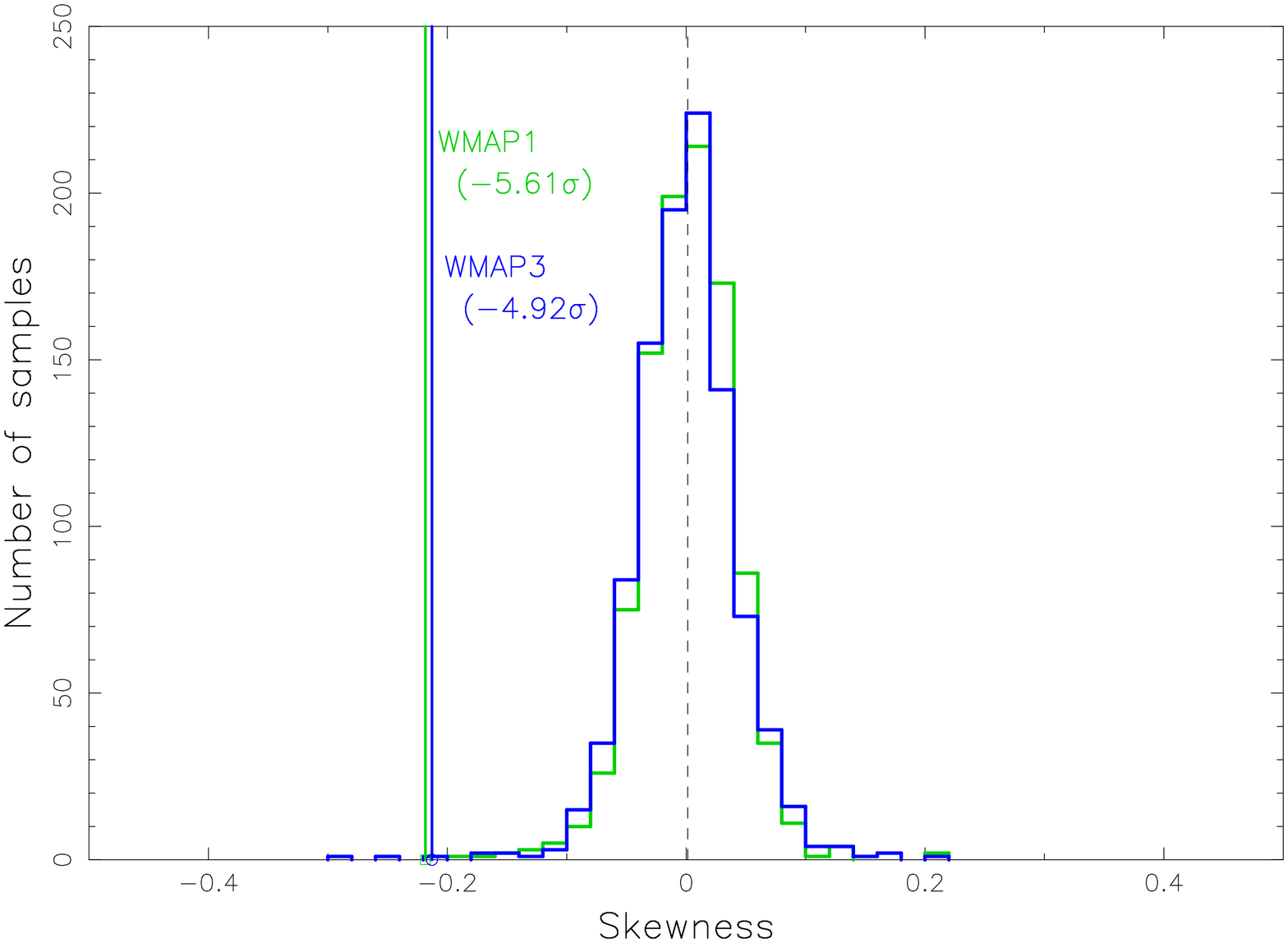}
\caption{Histograms of \morlet\ wavelet coefficient skewness \mbox{($\scale_{11}=550\arcmin$; $\eulerc=72^\circ$)} obtained from \ngsim\ Monte Carlo simulations.  Histograms are plotted for simulations in accordance with \wmap 1  (green) and \wmap 3 (blue) observations.
The observed statistics for the \wmap 1 and \wmap 3 maps are shown by the green and blue lines
  respectively.
The number of standard deviations these observations deviate from the mean of the appropriate set of simulations is also displayed.
}
\label{fig:hist}
\end{figure}

Next we consider in more detail the most significant deviation from Gaussianity on scale $\scale_{11}=550\arcmin$ and orientation $\eulerc=72^\circ$.  \Fig{\ref{fig:hist}} shows histograms of this particular statistic constructed from the \wmap 1 and \wmap 3 Monte Carlo simulations.  The measured statistic for the \wmap 1 and \wmap 3 data is also shown on the plot, with the number of standard deviations each observation deviates from the mean of the appropriate set of simulations.  
The distribution of this skewness statistic is not significantly altered between simulations that are consistent with \wmap1 or \wmap3 data. The observed statistics for the \wmap1 and \wmap3 data are similar but the slightly lower value for \wmap3 is now more apparent.

\begin{figure}
\centering
  \includegraphics[angle=-90,width=75mm]{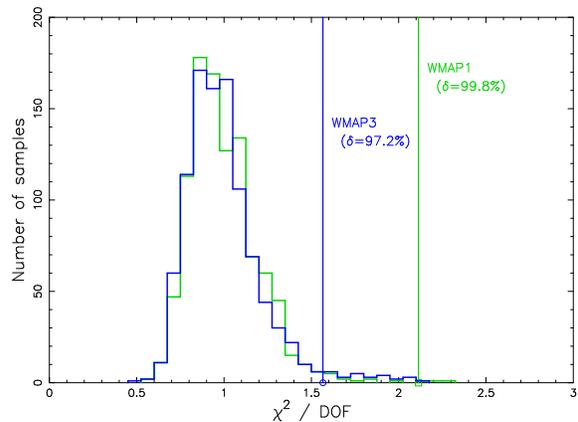}
\caption{Histograms of normalised $\chi^2$ test
  statistics computed from \morlet\ wavelet coefficient statistics obtained from \ngsim\ Monte Carlo simulations.  Histograms are plotted for simulations in accordance with \wmap 1  (green) and \wmap 3 (blue) observations.
The $\chi^2$ value computed for the \wmap 1 and \wmap 3 maps are shown by the green and blue lines
  respectively.  The significance of these observations, computed from the appropriate set of simulations, is also displayed.}
\label{fig:chi2}
\end{figure}

To quantify the statistical significance of the detected deviation from Gaussianity we consider two techniques.  The first technique involves comparing the deviation of the observed statistic to all statistics computed from the simulations.  This is a very conservative means of constructing significance levels.  The second technique involves performing a  $\chi^2$ test.  In both of these tests we relate the observation to all test statistics originally computed, \ie\ to both skewness and kurtosis statistics\footnote{Although we recognise the distinction between skewness and kurtosis, there is no reason to partition the set of test statistics into skewness and kurtosis subsets.  The full set of test statistics must be considered.}.  For a more thorough description of these techniques see \citet{mcewen:2005a}.
Searching through the \ngsim\ \wmap3 simulations, 51 maps have an equivalent or greater deviation that the \wmap3 data in any single test statistic computed for that map.  Using the very conservative first technique, the significance of the detection of non-Gaussianity in the \wmap 3 data may therefore be quoted at $94.9\pm0.7$\% (an expression for the $1\sigma$ errors quoted on significance levels is derived in \appn{\ref{appn:sig_errors}}).
The distribution of $\chi^2$ values obtained from the simulations is shown in \fig{\ref{fig:chi2}}.  The $\chi^2$ value obtained for the data is also shown on the plot.  The distribution of the $\chi^2$ values is not significantly altered between simulations that are consistent with \wmap1 or \wmap3 data.  The $\chi^2$ value computed for the data, however, is significantly lower for the \wmap3 data.  Computing the significance of the detection of non-Gaussianity directly from the $\chi^2$ distribution and observation, the significance of the detection of non-Gaussianity in the \wmap 3 data may be quoted at $97.2\pm0.5$\%.
Using both of the techniques outlined above the significance of the detection of non-Gaussianity made with the \wmap3 data is slightly lower than that made with the \wmap1 data.  Nevertheless, the non-Gaussian signal is still present at a significant level.

\begin{figure}
\centering
  \subfigure[\wmap 3 wavelet coefficients]{\includegraphics[width=70mm]{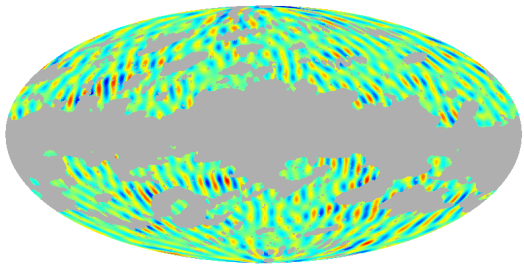}}
  \subfigure[\wmap 3 thresholded wavelet coefficients]{\includegraphics[width=70mm]{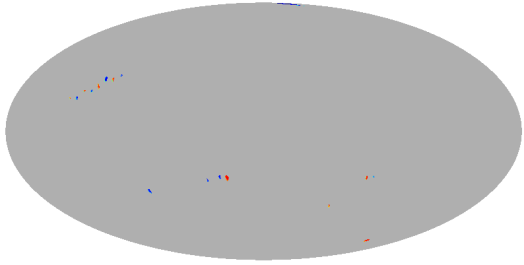}}
  \subfigure[\wmap 1 thresholded wavelet coefficients]{\includegraphics[width=70mm]{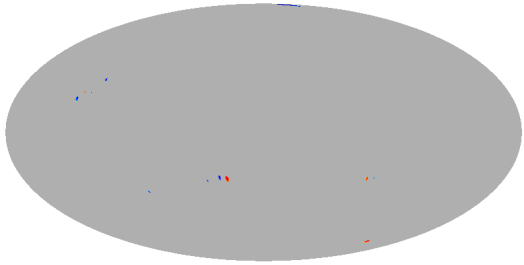}}
\caption{\Morlet\ spherical wavelet coefficient maps and \mbox{thresholded} versions \mbox{($\scale_{11}=550\arcmin$; $\eulerc=72^\circ$)}.
  To localise most likely deviations from Gaussianity on the sky, the
  coefficient map is thresholded
  so that only those coefficients above $3\sigma$ (in absolute value)
  remain.  All sky maps (here and subsequently) are illustrated in Galactic coordinates, with the Galactic center in the middle.
}
\label{fig:coeff}
\end{figure}

A wavelet analysis allows the spatial localisation of interesting signal characteristics.  The most pronounced deviations from Gaussianity in the \wmap\ data may therefore be localised on the sky.  The \morlet\ wavelet coefficients of the \wmap3 data corresponding to the most significant detection of non-Gaussianity on scale $\scale_{11}=550\arcmin$ and orientation $\eulerc=72^\circ$ are displayed in \fig{\ref{fig:coeff}}.  Thresholded wavelet coefficient maps for both the \wmap1 and \wmap3 data are also shown in order to localise the most pronounced deviations.  The regions localised in the \wmap1 and \wmap3 data are very similar, although the localised regions appear slightly more pronounced, in the sense that the peaks are larger, in the \wmap3 data.  To investigate the impact of localised regions on the initial detection of non-Gaussianity, the analysis is repeated with the \wmap3 localised regions excluded from the analysis.
The resulting skewness statistics are shown by the dashed line in \fig{\ref{fig:stats}}.  Interestingly, the highly significant detections of non-Gaussianity are eliminated when these localised regions are removed.

In our previous non-Gaussianity analysis \citep{mcewen:2005a} we also performed a preliminary noise analysis and found that noise was not atypical in the localised regions that we detect.  The localised regions have not changed in the \wmap3 data, hence we do not expect this finding to change.  In an additional work of ours \citep{mcewen:2005c} that investigated a Bianchi VII$_{\rm h}$ component as a possible source of non-Gaussianity -- which, incidentally, we found not to be the predominant source of non-Gaussianity -- we performed a preliminary analysis of foregrounds and systematics.  We concluded that foregrounds or systematics were not the likely source of the detected non-Gaussianity.  Again, we do not believe this finding to change in the \wmap3 data since both foregrounds and systematics are treated more thoroughly.


\section{Summary and discussion}
\label{sec:conclusions}

We have repeated on the \wmap3 data the directional spherical \morlet\ wavelet analysis used to make a significant detection of non-Gaussianity in the \wmap1 data \citep{mcewen:2005a}.  The non-Gaussian signal previously detected is indeed present in the \wmap3 data, although the significance of the detection is reduced.  Using our first very conservative method for constructing significance measures we find the significance of the detection of non-Gaussianity drops from $98.3\pm0.4$\% to $94.9\pm0.7$\%.  Using our second technique for constructing significance measures, which is based on a $\chi^2$ analysis, the significance of the detection drops from $99.3\pm0.3$\% to $97.2\pm0.5$\%.  We have no intuitive explanation for this drop in significance.  

\begin{figure}
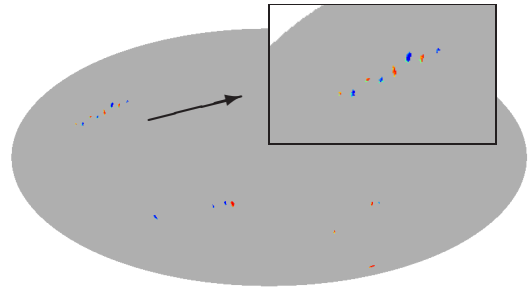


  \vspace{5mm}

  \begin{minipage}[t]{80mm}
    \vspace{0pt}
    \hspace{5mm}
    \includegraphics[width=70mm]{figures/wmap3_morlet_thres_ia11_ig02_display_sc}
  \end{minipage}\hspace{-40mm}
  \begin{minipage}[t]{25mm}
    \vspace{-3mm}
    \hspace{10mm}
    \frame{\includegraphics[bb= 50 210 230 320,width=30mm,clip]{figures/wmap3_morlet_thres_ia11_ig02_display_sc}}
  \end{minipage}

  \begin{minipage}[t]{20mm}
    \begin{picture}(0,0)(-70,-65)
      \thicklines
      \put(0,0){\vector(4,1){35}}
    \end{picture}
  \end{minipage}

\caption{Thresholded \morlet\ spherical wavelet coefficient map with localised regions inset \mbox{($\scale_{11}=550\arcmin$; $\eulerc=72^\circ$)}.}
\label{fig:coeff2}
\end{figure}

The most likely sources of non-Gaussianity were also localised on the sky.  We detect the same regions in the \wmap3 data as found in the \wmap1 data, although the localised regions extracted appear slightly more pronounced in the \wmap3 data.  When all localised regions are excluded from the analysis the data is consistent with Gaussianity.  An interesting structure is extracted in the
upper-left region of the thresholded maps (see \fig{\ref{fig:coeff2}}), \ie\ in the vicinity of Galactic coordinates $(l,b)=(120^\circ,25^\circ)$.
In a future work we intend to use optimal filters on the sphere, in conjunction with our fast \cswt\ analysis tool, to search for cosmic strings in the \cmb, a possible source of the non-Gaussianity that we have detected in both the \wmap1 and \wmap3 data.

\section*{Acknowledgements}

JDM thanks the Association of Commonwealth
Universities and the Cambridge Commonwealth Trust for the 
support of a Commonwealth (Cambridge) Scholarship.
Some of the results in this paper have been derived using the
\healpix\ package \citep{gorski:2005}.
We acknowledge the use of the Legacy Archive for Microwave Background
Data Analysis (\lambdaarch).  Support for \lambdaarch\ is provided by
the NASA Office of Space Science.


\appendix
\section{Errors on significance levels}
\label{appn:sig_errors}

In this appendix we derive the standard deviation of a significance level determined from Monte Carlo (MC) simulations.  Suppose we perform $n$ independent MC simulations.  Let $p$ denote the probability that an MC simulation chosen at random has a value for some test statistic that is larger that the corresponding value derived from the real data (hence $p$ is the underlying significance we attempt to estimate).  Choosing an MC simulation at random and defining whether it has a test statistic greater than that of the data thus corresponds to a Bernoulli trial with a probability of success equal to $p$.

Suppose we observe $x$ successes in the $n$ MC simulations.  The likelihood for $x$ is
\begin{displaymath}
{\rm Pr}(x|p) = {}^n C_{x} p^{x} (1-p)^{n-x}
\spcend .
\end{displaymath}
The maximum likelihood (ML) estimate $\hat{p}$ of $p$ is most easily given by maximiming the log-likelihood:
\begin{displaymath}
\left.
\frac{\partial {\rm ln} {\rm Pr}(x|p)}{\partial p} \right|_{p=\hat{p}} = 0
\;\;\;\;\;
\Rightarrow
\;\;\;\;\;
\hat{p} = \frac{x}{n} 
\spcend ,
\end{displaymath}
which recovers the intuitive result.
Approximating the shape of the likelihood near its peak by a Gaussian, we may approximate the standard deviation of $\hat{p}$ by
\begin{displaymath}
\sigma_{\hat{p}} = 
\left[\left.
-
\frac{\partial^2 {\rm ln} {\rm Pr}(x|p)}{\partial p^2} \right|_{p=\hat{p}}
\right]^{-1/2}
=
\sqrt{
\frac{x(n-x)}{n^3}
}
\spcend .
\end{displaymath}


\label{lastpage}

\end{document}